\documentclass[prb,twocolumn,showpacs,floatfix]{revtex4}
\usepackage{graphicx,color}
\usepackage{amssymb}
\usepackage{amsmath}
\usepackage{xspace}
\usepackage{url}

\let \Re \relax
\DeclareMathOperator{\Re}{Re}

\begin{document} 
\title{Slave-boson field fluctuation approach to the extended Falicov-Kimball model:\\ charge, orbital, and excitonic susceptibilities}

%
\author{B. Zenker$^1$, D. Ihle$^2$, F. X. Bronold$^1$, and H. Fehske$^{1,3}$}
\affiliation{$^1$Institut f{\"u}r Physik,
            Ernst-Moritz-Arndt-Universit{\"a}t Greifswald,
             D-17489 Greifswald, Germany \\
             $^2$Institut f{\"u}r Theoretische Physik, Universit{\"a}t Leipzig, D-04109 Leipzig, Germany\\
             $^3$School of Physics, University of New South Wales, 
Sydney NSW 2052, Australia}

\date{\today}
\begin{abstract}
Based on the $SO(2)$-invariant slave-boson scheme, the static charge, orbital, and excitonic susceptibilities in the extended Falicov-Kimball model are calculated. Analyzing the phase without long-range order we find instabilities towards charge order, orbital order, and the excitonic insulator (EI) phase. The instability towards the EI is in agreement with the saddle-point phase diagram. We also evaluate the dynamic excitonic susceptibility, which allows the investigation of uncondensed excitons. We find  qualitatively different features of the exciton dispersion at the semimetal-EI and at the semiconductor-EI transition supporting a crossover scenario between a BCS-type electron-hole condensation and a Bose-Einstein condensation of preformed bound electron-hole pairs.
\end{abstract}
\maketitle

\section{Introduction} 
At low temperatures electronic correlations can cause anomalies at the semimetal-semiconductor (SM-SC) transition.~\cite{HR68} Half a century ago, Mott~\cite{M61} argued that in a SM with a very low carrier density the Coulomb attraction between electrons and holes should lead to the spontaneous formation of electron-hole bound states (excitons), and the system would become insulating. Shortly afterwards, Knox~\cite{Kno63} noticed that a SC is unstable against the spontaneous formation of excitons if the exciton binding energy overcomes the gap energy separating valence and conduction band. Both arguments suggest a new distorted phase, an exciton condensate known as the excitonic insulator (EI), to be the crystal ground state.
The SM-EI transition is mathematically similar to the BCS theory of superconductivity, while the SC-EI transition can be treated as a Bose-Einstein condensation (BEC) of preformed excitons. Hence, the EI is discussed in view of a BCS-BEC crossover scenario in a solid.~\cite{CN82a, BF06,IPBBF08,PBF10}

Whilst theoretically predicted a long time ago,~\cite{C65} (for recent reviews see Ref.~\onlinecite{LEKMSS04}) no conclusive experimental proof of the existence of the EI has been achieved yet. However, there are a few promising candidates. In the mixed valence compound TmSe$_{0.45}$Te$_{0.55}$ detailed studies of the pressure-induced SC-SM transition suggest that excitons are created in a large number and condense below 20 K.~\cite{NW90} More recently, several transition-metal dichalcogenides were reported to exhibit an EI phase. Angle-resolved photo emission spectra (ARPES) measurements of Ta$_2$NiSe$_5$ traced back the extreme valence band top flattening at low temperature to an EI ground state.~\cite{WaEtAl09}  ARPES data of $1T$-TiSe$_2$ indicate that the EI is the driving force for the charge-density-wave (CDW) transition in this material.~\cite{CeEtAl07}

\begin{figure}
\includegraphics[width=0.95\linewidth]{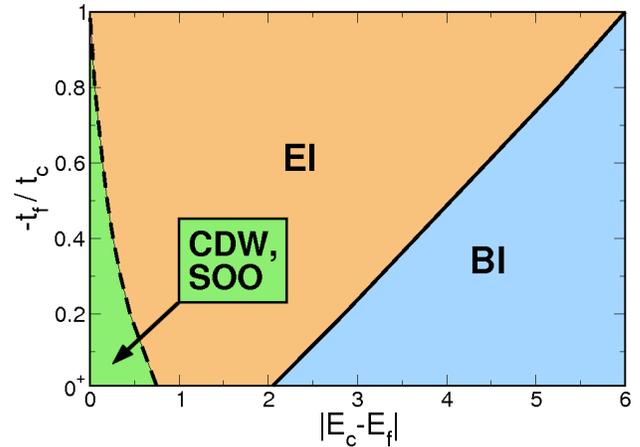}
\caption{(Color online) Hartree-Fock ground-state phase diagram of the EFKM in two dimensions for Coulomb strength $U=2$. The difference between CDW and SOO is explained in the text. The black solid line represents the second-order transition from an EI to a  BI (band insulator), 
the dashed line represents the first-order CDW/SOO-EI transition. The 
label $0^+$ emphasizes that the EI phase is present only for finite 
$f$-bandwidths.}
\label{fig1}
\end{figure}

From a theoretical point of view, the description of the EI with a Falicov-Kimball-type model seems promising. The original Falicov-Kimball (FKM) model~\cite{FK69} contains itinerant $c$-electrons (with bandcenter $E_c$ and hopping amplitude $t_c$) that interact via a local Coulomb repulsion $U$ with localized $f$-electrons (with energy level $E_f$), where the spin is neglected. Since the local $f$-electron number is strictly conserved in the FKM, $f$-$c$-coherence cannot be established.~\cite{SB88} One way to overcome this shortcoming is to include an $f$-$c$ hybridization.~\cite{KMM76} As shown in Refs.~\onlinecite{Ba02b, BGBL04}, the extension by a finite $f$-bandwidth also induces $f$-$c$ coherence. The model with a direct $f$-$f$ hopping (with hopping amplitude $t_f$) is called the extended Falicov-Kimball model (EFKM) and has previously been used to describe different properties of the EI phase.~\cite{Br08, IPBBF08, ZIBF10, PBF10}
The ground-state phase diagram of the EFKM was determined with a constraint path Monte Carlo (CPMC) technique for one and two dimensions (1D and 2D) in the strong~\cite{Ba02b} and intermediate coupling regime~\cite{BGBL04} as well as in the Hartree-Fock (HF) approximation for 2D,~\cite{Fa08} 3D,~\cite{Fa08} and infinite dimensions.~\cite{SC08} All approaches yield a qualitatively similar phase diagram.
 Figure~\ref{fig1} displays the HF ground-state phase diagram for $U=2$ in 2D, depicting the generic situation. It was shown previously that Fig.~\ref{fig1} agrees with the CPMC data even quantitatively.~\cite{Fa08}
Besides full $c$- and full $f$-band insulator (BI) regions, the EFKM ground-state phase diagram exhibits three symmetry broken phases: the EI, a CDW, and a staggered orbital order (SOO). The EI is characterized by a nonvanishing average $\langle c^\dagger f^{} \rangle$. 
The CDW is described by a periodic modulation in the total electron density comprising both $f$- and $c$-electrons. The SOO is characterized by a periodic modulation in the difference between the $f$-electron and the $c$-electron density, which may be accompanied by a CDW.
The SOO (CDW) establishes the ground state for the symmetric case ($E_f=E_c$) for all ratios of $-t_f/t_c$ (for the CDW the point $|t_f|=|t_c|$ has to be excluded, see below).  These phases are rapidly suppressed in favor of the EI if $E_f\neq E_c$.   Between the uniform EI phase and the CDW or SOO phase there is a first-order phase transition. The EI-BI transition is of second order. For $t_f=0$ the FKM is recovered, and the EI phase cannot be realized.

For the investigation of electron correlation effects the Gutzwiller approximation~\cite{Gu63} is an established technique. Kotliar and Ruckenstein introduced a scalar slave-boson (SB) scheme which reproduces the Gutzwiller solution of the Hubbard model as a saddle-point.~\cite{KR86} A manifestly spin-rotation invariant form of the SB representation has been worked out for the Hubbard model~\cite{LWH89} and for multiband Hubbard models.~\cite{LGKP07} 
We have developed an $SO(2)$-invariant SB approach for the EFKM (Ref.~\onlinecite{ZIBF10}) that reproduces the HF result for the EI phase boundary at $T=0$, but leads to a substantial reduction of the critical temperature. 
It is the aim of this work to include Gaussian fluctuations around the saddle-point~\cite{La90, LSW91, DFKTI94} in the $SO(2)$-invariant SB scheme at zero and finite temperature. This offers an opportunity to calculate susceptibilities for investigating instabilities against long-range ordered phases and the formation of excitons.

The paper is organized as follows. In Sec.~\ref{Theory} the model Hamiltonian and the SB scheme are introduced. Moreover the saddle-point approximation is given and the calculation of response functions within the SB scheme is explained. 
In Sec.~\ref{numerics} we present numerical results for the instabilities toward the CDW, the SOO, and the EI phase. Finally we investigate the formation of excitons in the phase without long-range order. Section~\ref{summary} summarizes our results.

\section{Theory}\label{Theory}
\subsection{Model Hamiltonian}\label{Hamiltonian}
Expressing the orbital flavor by a pseudospin variable $\sigma=\uparrow,\downarrow$, where $c_{i\uparrow}^{(\dagger)}\equiv f_i^{(\dagger)}$ and $c_{i\downarrow}^{(\dagger)}\equiv c_i^{(\dagger)}$, the EFKM can be written as an asymmetric Hubbard model,
\begin{equation}
H=\sum_{i,\sigma} (E_\sigma-\mu) c_{i\sigma}^\dagger c_{i\sigma}^{ } - \sum_{\langle i,j\rangle,\sigma} t_\sigma c_{i\sigma}^\dagger c_{j\sigma}^{} + U\sum_i n_{i\uparrow} n_{i\downarrow}\;, \label{EFKM}
\end{equation}
where $c_{i\sigma}^{(\dagger)}$ annihilates (creates) a $\sigma$-band electron at the Wannier site $i$ and $n_{i\sigma}^{ }=c_{i\sigma}^\dagger c_{i\sigma}^{ }$ is the corresponding number operator.  $E_\sigma$ denotes the bandcenter of the $\sigma$-electron band, $\mu$ gives the chemical potential, $t_\sigma$ is the hopping amplitude, and $U$ measures the Coulomb interaction strength. In what follows we consider $E_\downarrow=0$, $E_\uparrow<0$, $t_\downarrow=1$, and $t_\uparrow<0$. All energies are measured in units of $t_\downarrow$. We restrict ourselves to  $t_\downarrow t_\uparrow<0$, i.e., the valence band top and the conduction band minimum are located at the Brillouin zone center. Moreover we exclusively investigate the half-filled band case, i.e., $\tfrac{1}{N} \sum_{i,\sigma} \langle n_{i\sigma} \rangle =1$, where $N$ is the number of lattice sites.

\subsection{Slave-boson functional integral representation}\label{SB-representation}
Following Refs.~\onlinecite{LWH89, ZIBF10} the Hilbert space is enlarged by introducing auxiliary bosons: $e_i^{(\dagger)}$, related to an empty site, $d_i^{(\dagger)}$, related to a doubly occupied site, and $\underline{p}_i^{(\dagger)}$, related to a singly occupied site,
\begin{align}
|0_i\rangle\;\;\rightarrow &\;\;e_i^\dagger |\mathrm{vac}\rangle\,,\\
|2_i\rangle\;\;\rightarrow &\;\;\tilde{c}_{i\uparrow}^\dagger 
\tilde{c}_{i\downarrow}^\dagger d_i^\dagger|\mathrm{vac}\rangle\,,\\
|\sigma_i \rangle\;\;\rightarrow &\;\; \sum_{\rho}
\tilde{c}_{i\rho}^\dagger p_{i\rho\sigma}^\dagger |\mathrm{vac}\rangle\,,
\end{align}
where $ |\mathrm{vac}\rangle$ means the vacuum state.
The matrix operator $\underline{p}_i^{(\dagger)}$ is given as
\begin{equation}
\underline{p}_i^{(\dagger)}=\frac{1}{2}\sum_{\mu} 
\underline{\tau}_\mu p_{i\mu}^{(\dagger)}
=\frac{1}{2}\begin{pmatrix} 
p_{i0}^{(\dagger)} +p_{iz}^{(\dagger)} & p_{ix}^{(\dagger)} -ip_{iy}^{(\dagger)}
\\[0.1cm] p_{ix}^{(\dagger)} +ip_{iy}^{(\dagger)} & p_{i0}^{(\dagger)}-p_{iz}^{(\dagger)} 
\end{pmatrix}\,,
\end{equation}
where $\mu=0,x,y,z$. $\underline{\tau}_0$ denotes the unit matrix and $\vec{\underline{\tau}}=(\underline{\tau}_x, \underline{\tau}_y, \underline{\tau}_z)^{\rm T}$ is the vector of the Pauli matrices. 
The fermionic degrees of freedom are captured by the pseudofermions $ \tilde{\mathbf{c}}^\dagger_i=(\tilde{c}_{i\uparrow}^\dagger , \tilde{c}_{i\downarrow}^\dagger)$ and $ \tilde{\mathbf{c}}^{}_i=(\tilde{c}_{i\uparrow}^{} , \tilde{c}_{i\downarrow}^{})^{\rm T}$.

Unphysical states of the extended fermion-boson Fock space are excluded by two sets of local constraints,
\begin{eqnarray}
C^{(1)}_i &=& e^\dagger_i e_i^{}\; 
	+ 2\, \mbox{Tr}\,
{\underline{p}}^\dagger_i{\underline{p}}^{}_i \,+ d^\dagger_id_i^{}\,- 1\,  =0\,,\label{c1}\\
\underline{C}^{(2)}_i &=& \tilde{\mathbf{c}}^{}_i \tilde{\mathbf{c}}^\dagger_i +
	2 \, {\underline{p}}^\dagger_i{\underline{p}}^{}_i
        + d^\dagger_id_i^{}\,\underline{\tau}_0 -\underline{\tau}^{}_0 =0\,.
\label{c2}
\end{eqnarray}

Since the bosonic occupation number of one site is coupled to the fermionic occupation, the bosons have to change simultaneously when an electron is created or annihilated. This is achieved by introducing the bosonic hopping operator $\underline{z}_i$,
\begin{equation}
c_{i\sigma} = \sum_\rho z_{i\sigma \rho} \tilde{c}_{i\rho} \label{bosonHop1} \;.
\end{equation}
The choice of $\underline{z}_i$ is not unique. We choose~\cite{LWH89, ZIBF10} 
\begin{equation}
{\underline{z}_i} = {\underline{L}_i}e^\dagger_i
        {{M}_i}{\underline{p}^{}_i} {\underline{N}_i}
        + {\underline{L}_i} {\tilde{\underline{p}}^{\dagger}_i}
        {{M}_i} d^{}_i {\underline{N}_i}\;
\label{E_zsri}
\end{equation}
\hspace*{1cm}with
\begin{eqnarray}
        {\underline{L}}^{}_i &=& [( 1 - d_i^\dagger d^{}_i )\underline{\tau}_0 - 
        2{\underline{p}}^\dagger_i{\underline{p}}^{}_i ]^{-1/2}\;,\\
        {\underline{N}_i} &=& [( 1 - e_i^\dagger e^{}_i )\underline{\tau}_0 - 
        2{\underline{\tilde{p}}}^\dagger_i{\underline{\tilde{p}}}^{}_i
        ]^{-1/2}\;,\\
        M^{}_i &=& [1 + e^\dagger_i e^{}_i + d^\dagger_i d^{}_i 
        + 2\, \mathrm{Tr}\, {\underline{p}}^\dagger_i{\underline{p}}^{}_i]^{1/2}\:,
\label{LMR_d}
\end{eqnarray}
and \mbox{$\tilde{p}^{\left(\dagger\right)}_{i\rho\rho '} =
\rho\rho ' p^{\left(\dagger\right)}_{i-\rho' -\rho}\;$}, which guarantees the correct free-fermion result on the mean-field level.
The Hubbard interaction term is bosonized via $n_{i\uparrow} n_{i\downarrow} \rightarrow d_i^\dagger d_i^{}$.

The resulting coupled fermion-boson system is evaluated within a functional integral representation. Then, the bosons become complex fields and the fermions are given by complex Grassmann fields. The Lagrange multipliers $\lambda_i^{(1)}$, $\lambda_{i0}^{(2)}$, $\lambda_{ix}^{(2)}$, $\lambda_{iy}^{(2)}$, and $\lambda_{iz}^{(2)}$ are introduced to enforce the constraints~\eqref{c1} and \eqref{c2}. Exploiting the gauge symmetry of the action and transforming the Lagrange multipliers into real time-dependent Bose fields we can remove the phases of $p_{i0}$, $p_{iz},$ and $e_i$. Using the Grassmann integration formula, we obtain the grand canonical partition function given by a functional integral over Bose fields only,
\begin{align}	
Z=\int &D[e] D [p^{}_0] D [p^{\ast}_x,p^{}_x] D [p^{\ast}_y,p^{}_y] D [p^{}_z] D[d^\ast,d]  \nonumber \\
&D[\lambda^{(1)}] D[\lambda^{(2)}_0] D[\vec{\lambda}^{(2)}] \,\mathrm{e}^{-S}
\label{E_zusta2}
\end{align}
with the effective bosonic action 
\begin{align}
	S =& \int\limits_0^\beta d\tau \bigg\{\sum_i
	\Big[- \lambda^{(1)}_i+
	\lambda^{(1)}_i e^2_i + \sum_\mu ( \lambda^{(1)}_i -
	\lambda^{(2)}_{i0} ) |p_{i\mu}|^2 \nonumber \\
	&\qquad\quad\qquad- p^{ }_{i0}(\vec{p}^{\,\ast }_i+\vec{p}^{ }_i)
        \vec{\lambda}^{(2)}_{i} 
	-i\vec{\lambda}^{(2)}_i (\vec{p}^{\,\ast}_i\times \vec{p}^{ }_i)
          \nonumber \\[0.1cm] 
        &\qquad\quad\qquad+(\lambda_i^{(1)} + U-2\lambda_{i0}^{(2)}) |d_{i}|^2 
         \nonumber \\
	&\qquad\quad\qquad+ p^\ast_{ix}\partial_\tau p_{ix}+p^\ast_{iy}\partial_\tau p_{iy}+d_i^\ast \partial_\tau d_i^{}
	\Big]\bigg\}\nonumber \\ 
	       &\;\;- \mbox{Tr}\, 
	\ln\Big\{-G^{-1}_{\langle ij\rangle,\rho \rho '}(\tau, \tau ')\Big\}\,, 
\label{E_seff}
\end{align}
where $\vec{p}_i=(p_{ix},p_{iy},p_{iz})$ and $\vec{\lambda}_i^{(2)}=(\lambda_{ix}^{(2)}, \lambda_{iy}^{(2)},\lambda_{iz}^{(2)})$. The inverse Green propagator is given by
\begin{align}
	G^{-1}_{\langle ij\rangle,\rho \rho '}(\tau, \tau ') =& 
	\Big[
	\big( -\partial^{ }_\tau + \mu - \lambda^{(2)}_{i0}\big)\delta^{ }_{\rho \rho '}
	\nonumber\\
        &- \tfrac{E_\uparrow}{2} (\underline{\tau}_0
        +\underline{\tau}_z)_{\rho \rho'}-\vec{\lambda}^{(2)}_{i} 
	\vec{\tau}_{\rho \rho '} 
	\Big]\delta^{ }_{i j}\,\delta (\tau - \tau ')\nonumber\\
	&+
	(\underline{z}^\ast_i\,\underline{t}\,\underline{z}^{}_j)^{}_{\rho \rho ', 
	\tau \tau '}(1-\delta^{ }_{i j})\,,
\label{E_Gij}
\end{align}
where $\underline{t}=\begin{pmatrix} t_\uparrow & 0 \\ 0 & t_\downarrow\end{pmatrix}$.
The trace in Eq.~(\ref{E_seff}) extends over time, space, and spin variables. For the half-filled band case Eqs.~\eqref{E_seff} and \eqref{E_Gij} are an exact representation of the partition function of the EFKM. One obtains $\underline{z}_i=z_i \underline{\tau}_0$.

\subsection{Saddle-point approximation} \label{saddle-point}
To proceed we approximate all bosonic fields
by their time-averaged values (static approximation), i.e., the bosonic fields are taken to be real. Moreover, we look for uniform solutions, that is, the Bose fields are taken to be independent of the lattice site.

We restrict ourselves to the phase without long-range order, which we denote as paraphase.
The saddle-point equations for the paraphase ($p_{x}=p_{y}=\lambda_{x}^{(2)}=\lambda_{y}^{(2)}=0$) are
\begin{eqnarray}
p_0 p_z &=& \frac{1}{2}\left( n_\uparrow-n_\downarrow \right) \;, \label{p0pz} \\
p_0^2 &=& \frac{1}{2} +\sqrt{n_\uparrow n_\downarrow (1-z^2)} \;, \label{p02} \\
d^2 &=& \frac{1}{2 z^2} \biggl[ z^2 \left( 2-p_0^2-p_z^2\right) + 2p_0^2\nonumber \\
&& -2p_0 \sqrt{z^2\left(2-p_0^2-p_z^2\right)+z^4p_z^2+p_0^2}\biggr]\;,\label{d} \\
\lambda_z^{(2)} &=& -\frac{p_z}{p_0}\left(\frac{z^2}{2d^2}-\frac{1}{p_0^2-p_z^2}\right) z^2 \epsilon(0) \;,\label{lambda_z} \\
U&=&-\frac{2d^2-p_0^2+z^2p_z^2}{p_0^2d^2} z^2 \epsilon(0)-2\lambda_z^{(2)} \frac{p_z}{p_0}\;,\label{corrEqn} \\
n_\sigma &=& \frac{1}{N}\sum_{\bf k} n_{{\bf k}\sigma} \label{occupation_number} \;,\\
\epsilon(0) &=& \frac{1}{N} \sum_{\bf k} \left( t_\uparrow \gamma_{\bf k} n_{{\bf k}\uparrow} + t_\downarrow \gamma_{\bf k} n_{{\bf k}\downarrow}\right) \;,\label{epsilon}
\end{eqnarray}
where 
\begin{eqnarray}
n_{{\bf k}\sigma} &=& [\exp( \beta E_{{\bf k}\sigma})+1]^{-1} \;, \label{n_k} \\
E_{{\bf k}\sigma}& =& E_\sigma +\sigma \lambda_z^{(2)}-\tilde\mu-z^2 t_\sigma \gamma_{\bf k} \;, \label{Ek_updown} \\ 
\tilde \mu &=&\mu-\lambda_0^{(2)} \;. \label{mu_tilde}
\end{eqnarray}
On a D-dimensional hypercubic lattice, $ \gamma_{\bf k}=2\sum_{l=1}^D \cos k_l$. 
The chemical potential is determined by the condition
\begin{equation}
\frac{1}{N}\sum_{{\bf k},\sigma} n_{{\bf k}\sigma}=1\,.
\label{mu2}
\end{equation}
  
The quasiparticle gap $E_g$ indicates the splitting of the $\uparrow$- and $\downarrow$-band (in the paraphase), which is caused by the correlation-induced quasiparticle bandshift $\lambda_z^{(2)}$. For a D-dimensional hypercubic lattice, $E_g$ is given by
\begin{equation}
E_g=|E_\uparrow|+|2\lambda_z^{(2)}|-2Dz^2(|t_\uparrow|+|t_\downarrow|) \;. \label{Hartree}
\end{equation}
For a SM, $E_g\leq0$ and for a SC, $E_g>0$.

We obtain the EI phase boundary by solving the SB gap equation,
\begin{equation}
 1= \frac{1}{p_0 p_z} \lambda_z^{(2)}  \frac{1}{N} \sum_{\bf k} \frac{n_{{\bf k}\uparrow} -n_{{\bf k}\downarrow}}{E_{{\bf k}\uparrow}-E_{{\bf k}\downarrow}} \;, \label{SBGapEq}
\end{equation}
resulting from Eqs.~(63) and (65) of Ref.~\onlinecite{ZIBF10}.

The gap equation~\eqref{SBGapEq} captures both the BCS and the BEC situation, but it cannot discriminate between them. To this end, we follow an idea from Ihle et al. (Ref.~\onlinecite{IPBBF08}) and investigate the excitonic susceptibility in the paraphase.

\subsection{Gaussian fluctuations}\label{Gaussian-fluct}
In order to study response functions, we take into account Gaussian fluctuations around the saddle point for the paraphase, that is, $\Phi_{ia} = \bar\Phi_a + \delta\Phi_{ia}$, where $\Phi_{ia}=\{e_i,p_{i0},p_{ix},p_{iy},p_{iz},d_i,\lambda_i^{(1)},\lambda_{i0}^{(2)}, \lambda_{ix}^{(2)}, \lambda_{iy}^{(2)}, \lambda_{iz}^{(2)}\}$. Then, the action is given by
\begin{equation}
S = \bar{S} + \sum_{q,a,b} \delta\Phi_a(-q) {\cal S}_{ab}(q) \delta\Phi_b(q) \;,
\label{fluct_expansion}
\end{equation}
where the bar denotes the saddle-point value. 

In order to achieve comparability with the saddle-point results, we start the fluctuation calculation from the same level of approximation as for the saddle-point calculation, i.e., we first perform the static approximation and consider only the fluctuations of the 11 real-valued fields $\Phi_{ia}$.

The fluctuation matrix can be calculated according to 
\begin{align}
{\cal S}&_{ab}(q,q')   \nonumber \\
=&\; \frac{1}{2N\beta} \sum_{R_i,R_j} e^{-iqR_i} \frac{\partial^2 S}{\partial \Phi_{ia} \partial \Phi_{jb}}\Bigg|_{\tiny\begin{matrix}\Phi_i=\Phi_j=\bar{\Phi}\\\tau=\tau'\end{matrix}} e^{-iq'R_j}\nonumber\\=&\;{\cal S}_{ab}(q)\delta_{q,-q'}\;.
\label{fluct_matrix1}
\end{align}
Here, we use the shorthand notation ${R}_i=({\bf R}_i, \tau)$ and ${q}=({\bf q}, \omega_m)$, where $\tau$ is the imaginary time, $\omega_m=2\pi m/\beta$ denote the bosonic Matsubara frequencies, ${\bf R}_i$ is the position vector, and ${\bf q}$ is the wave vector. 

The response functions can be expressed in terms of the SB field fluctuations using the local constraints \eqref{c1} and  \eqref{c2}.
The charge susceptibility reads
\begin{eqnarray}
\chi_c(q) &=& \langle \delta\left[n_{\uparrow}(-q) + n_\downarrow(-q)\right] \delta\left[n_{\uparrow}(q) + n_\downarrow(q)\right] \rangle \nonumber \\
&=& 4 \left( e^2 \langle \delta e(-q) \delta e(q) \rangle - 2ed \langle \delta e(-q) \delta d(q)\rangle \right. \nonumber \\
&&\left.+ d^2 \langle \delta d(-q) \delta d(q) \rangle \right) \label{charge_sus}\;.
\end{eqnarray}
The orbital susceptibility is given by
\begin{eqnarray}
\chi_o(q) &=& \langle \delta\left[n_{\uparrow}(-q) - n_\downarrow(-q)\right] \delta\left[n_{\uparrow}(q) - n_\downarrow(q)\right] \rangle \nonumber \\
&=&4 \left( p_z^2 \langle \delta p_0(-q) \delta p_0(q) \rangle + 2p_z p_0\langle \delta p_0(-q) \delta p_z(q) \rangle \right.\nonumber \\
&&\left.+ p_0^2 \langle \delta p_z(-q) \delta p_z(q) \rangle \right)\label{spin_sus}\;.
\end{eqnarray}
Considering the creation operator of an onsite electron-hole pair~\cite{IPBBF08}  
\begin{equation}
b_i^\dagger = c_{i\downarrow}^\dagger c_{i\uparrow}^{\,} \;, \hspace{0.5cm}
b_{\bf q}^\dagger = \frac{1}{\sqrt{N}}\sum_{\bf k} c_{{\bf k}+{\bf q}\downarrow}^\dagger c_{{\bf k}\uparrow}^{} \;,\label{X-op2}
\end{equation}
the electron-hole susceptibility, hereafter denoted as excitonic susceptibility, is given by
\begin{eqnarray}
\chi_X(q) &=& \langle \delta b_{\bf q}^{} \,\delta b_{\bf q}^\dagger \rangle   \nonumber \\
&=& p_0^2\left[ \langle \delta p_x(-q)\,\delta p_x(q)\rangle  +  \langle \delta p_y(-q)\,\delta p_y(q)\rangle  \right.\nonumber\\
&&\left.   -i\langle \delta p_y(-q)\,\delta p_x(q)\rangle+i\langle \delta p_x(-q)\,\delta p_y(q)\rangle \right] \;. \nonumber \\\label{X-prop2}
\end{eqnarray}

The correlation functions may be expressed as functional integrals over Bose fields:
\begin{equation}
\langle \delta\Phi_a(-q) \delta\Phi_b(q)\rangle = \frac{1}{Z} \int D[\Phi] \; \delta\Phi_a(-q) \delta\Phi_b(q)\; e^{-{\cal S}(q)}\;.\label{CorrFuncGen}
\end{equation}
Hence, the correlation functions are related to the inverse fluctuation matrix by
\begin{equation}
\langle \delta\Phi_a(-q) \delta\Phi_b(q)\rangle = \frac{1}{2} {\cal S}_{ab}^{-1}(q)\;. \label{CorFluct}
\end{equation}

It turns out that for the paraphase the $11\times 11$ fluctuation matrix decomposes into a $7\times 7$ matrix containing the charge fluctuations ($\delta e$, $\delta p_0$, $\delta d$, $\delta \lambda^{(1)}$, $\delta \lambda_0^{(2)}$) and the orbital fluctuations ($\delta p_z$, $\delta \lambda_z^{(2)}$) and into a $4\times 4$ matrix containing the electron-hole pair fluctuations  ($\delta p_x$, $\delta \lambda_x^{(2)}$, $\delta p_y$, $\delta \lambda_y^{(2)}$). The SB fields are obtained by solving the saddle-point equations~\eqref{p0pz}--\eqref{epsilon} self-consistently.

The description of the CDW and SOO requires the inclusion of inhomogeneous solutions with a periodic modulation in the densities,
$\langle n_{i\sigma} \rangle = n_\sigma + \delta_\sigma \cos({\bf QR}_i)$,
where the order vector in 3D is given by ${\bf Q}=(\pi,\pi,\pi)$.
The CDW and SOO order parameters are
$\delta_{\rm CDW} =\tfrac{1}{2}(\delta_\uparrow+\delta_\downarrow)$ and 
$\delta_{\rm SOO} = \tfrac{1}{2}(\delta_\uparrow-\delta_\downarrow)$, respectively.~\cite{ZFB10}
If $|\delta_\uparrow|\neq |\delta_\downarrow|$, the CDW and SOO describe the same symmetry broken state.
We can investigate the formation of both phases without generalizing the SB formalism to a bipartite lattice by calculating the static ($\omega=0$) charge and orbital susceptibility with order vector ${\bf q}={\bf Q}$, given by
\begin{eqnarray}
\chi_c &=&\chi_c({\bf Q},0)\nonumber \\
&=& 2\left[e^2({\cal S}^{-1})_{ee}+d^2({\cal S}^{-1})_{dd}-2ed({\cal S}^{-1})_{ed}\right]\;,\label{charge_sus2} \\
\chi_o &=& \chi_o({\bf Q},0) \nonumber \\
&=& 2\left[p_z^2({\cal S}^{-1})_{p_0p_0}+p_0^2({\cal S}^{-1})_{p_zp_z}-2p_0p_z({\cal S}^{-1})_{p_0p_z}\right]\;.\nonumber \\\label{spin_sus2}
\end{eqnarray}
The inversion of the $7\times7$ matrix is performed numerically.

After analytic continuation ($i\omega_m\rightarrow \omega +i0^+$) the excitonic susceptibility \eqref{X-prop2} yields
\begin{equation}
\chi_X({\bf q},\omega) = \frac{\chi_X^{(0)}({\bf q},\omega)}{-\tfrac{{\cal S}_{p_x p_x}}{p_0^2}\, \chi_X^{(0)}({\bf q},\omega) +1 } \;, \label{X-prop3}
\end{equation}
with 
\begin{equation}
\chi_X^{(0)}({\bf q},\omega)=\frac{1}{N} \sum_{\bf k} \frac{n_{{\bf k}\uparrow}-n_{{\bf k}+{\bf q}\downarrow} }{\omega + E_{{\bf k}\uparrow}-E_{{\bf k}+{\bf q} \downarrow} } \; \label{G0}
\end{equation}
and
\begin{equation}
{\cal S}_{p_x p_x}=\left( \frac{1}{p_0^2} - \frac{1}{2} \frac{p_0^2-p_z^2}{p_0^2d^2} z^2 \right) z^2\,\epsilon(0) + \frac{p_z}{p_0} \lambda_z^{(2)} \;. \label{S_pxpx} 
\end{equation}
For the BI at $T=0$ the random phase approximation result~\cite{IPBBF08} is recovered, $-\frac{S_{p_x p_x}}{p_0^2}= U$.

To determine the EI phase we compute the static excitonic susceptibility $\chi_X({\bf q},0)$. The direct band gap situation gives the order vector of the EI phase as ${\bf q}=0$. Using Eq.~\eqref{lambda_z} the fluctuation matrix element ${\cal S}_{p_x p_x}$ [Eq.~\eqref{S_pxpx}] reduces to
\begin{equation}
{\cal S}_{p_x p_x} = \frac{p_0}{p_z} \lambda_z^{(2)} \;. \label{S_pxpx_gap}
\end{equation}
It is easy to see that the condition for the divergence of $\chi_X(0,0)$  equates to the gap equation~\eqref{SBGapEq}.

The poles of $\Re\chi_X^{(0)}({\bf q},\omega)$ [Eq.~\eqref{G0}] give the continuum of electron-hole excitations, i.e., $\omega_k({\bf q})=E_{{\bf k}+{\bf q} \downarrow}-E_{{\bf k}\uparrow}$. Excitonic pairing of electrons and holes is described by the pole of $\Re\chi_X({\bf q},\omega)$ [Eq.~\eqref{X-prop3}] outside the electron-hole continuum,~\cite{IPBBF08}  i.e., by
\begin{equation}
\Re\chi_X^{(0)}({\bf q},\omega_X) = \frac{p_0^2}{S_{p_x p_x} } \;, \label{X-cond}
\end{equation}
with
\begin{equation}
0 < \omega_X({\bf q}) < \omega_C({\bf q}) \;, \label{X-cond_energy}
\end{equation}
where $\omega_C({\bf q})=\omega_k({\bf q})|_{\rm min}$.
The exciton binding energy is given by 
\begin{equation}
E_X^B ({\bf q}) = \omega_C({\bf q}) - \omega_X({\bf q}) \;. \label{X-bind}
\end{equation}
We want to emphasize that $\omega_X$, $\omega_C$ and $E_X^B$ are explicitly ${\bf q}$-dependent in contrast to Ref.~\onlinecite{IPBBF08}, where only excitons with ${\bf q}=0$ are considered, and Ref.~\onlinecite{BF06}, where the exciton binding energy is assumed to be independent of ${\bf q}$.

\section{Numerical results}\label{numerics}
\subsection{Instabilities against CDW and SOO}\label{CDW-SOO}
To obtain results for the 3D EFKM we transform the ${\bf k}$-summation into an energy integral using the tight-binding density of states (DOS) for a simple cubic lattice.

From the charge and orbital susceptibility we derive information about the CDW and SOO formation, respectively. 
\begin{figure}[b]
\includegraphics[width=0.95\linewidth]{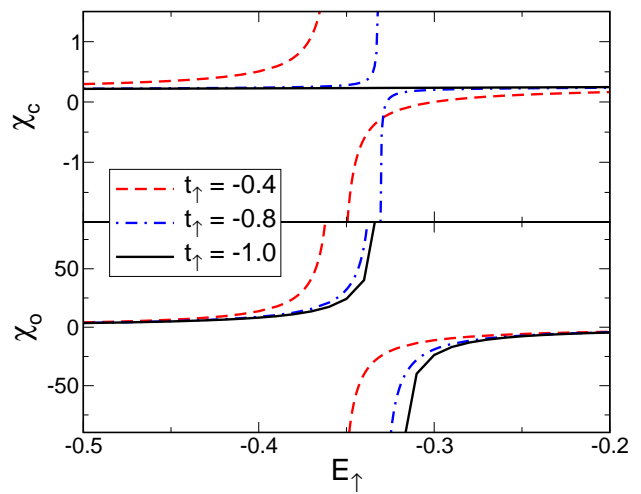}
\caption{(Color online) Static charge and orbital susceptibility of the 3D EFKM for $T=0$ and $U=4$ as a function of $E_\uparrow$.}
\label{fig2}
\end{figure}
For  asymmetric bands ($|t_\uparrow|\neq |t_\downarrow|$) the charge and orbital susceptibility diverge at the same critical $E_\uparrow$, as shown in Fig.~\ref{fig2}, implying $|\delta_\uparrow|\neq |\delta_\downarrow|$. The analogy between CDW and SOO vanishes if the bandwidths are equal, as can be seen for $t_\uparrow=-1.0$ in Fig.~\ref{fig2}. In this case, the orbital susceptibility diverges contrary to the charge susceptibility, thus, a CDW will not develop and $\delta_\uparrow=-\delta_\downarrow$. We conclude that the density inhomogeneity $\delta_\sigma$ is largely affected by the bandwidth.

\begin{figure}[h]
\includegraphics[width=0.95\linewidth]{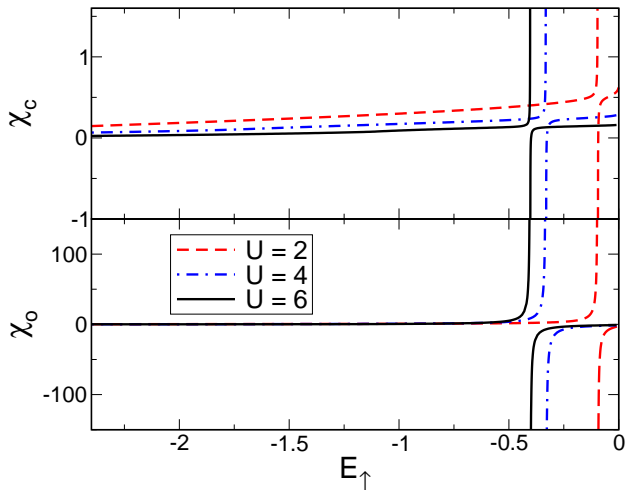}
\caption{(Color online) Static charge and orbital susceptibility of the 3D EFKM for $T=0$ and $t_\uparrow=-0.8$ as a function of $E_\uparrow$.}
\label{fig3}
\end{figure}
Figure~\ref{fig3} shows $\chi_o$ and $\chi_c$ for $t_\uparrow=-0.8$. The susceptibilities diverge at the same critical $E_\uparrow$. With increasing strength of the Coulomb interaction the critical $|E_\uparrow|$ for CDW (SOO) formation increases, because for a larger interaction the charge (orbital) order becomes more favorable. Figure~\ref{fig3} clearly shows that the CDW and SOO region is confined close to the symmetric case $E_\uparrow=0$. 

For small band splitting either the CDW (SOO) or the EI, separated by a first-order phase transition, can be realized, and one has to compare the free energies to identify the true ground state. Hence, to determine  the SB ground-state phase diagram (analogous to the HF case shown in Fig.~\ref{fig1}) the generalization of the saddle-point equations to a bipartite lattice is inevitable, which is beyond the scope of this work.
To investigate the EI in the following, we choose the band-structure parameters $E_\uparrow=-2.4$ and $t_\uparrow=-0.8$, where a CDW (SOO) is not realized (see Fig~\ref{fig3}).

\subsection{Instability against EI}\label{EI}
\begin{figure}[h]
\includegraphics[width=0.95\linewidth]{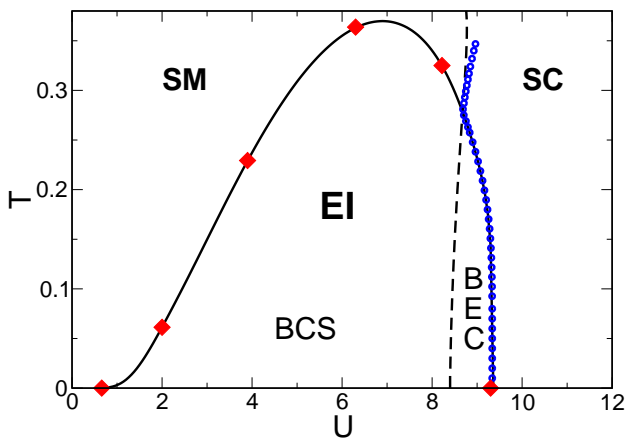}
\caption{(Color online) EI phase boundary (black solid line) of the 3D EFKM in the $U$-$T$ plane. The red diamonds give the poles of the uniform static excitonic susceptibility  for some fixed values of $U$. The blue circles give the critical $U_X$ for exciton formation with center-of-mass momentum ${\bf q}=0$. The black dashed line separates the SM and the SC phase. Inside the EI phase we suggest that the black dashed line gives a crude estimate for the BCS-BEC crossover region (see text). The band-structure parameters are $E_\uparrow=-2.4$ and $t_\uparrow=-0.8$.}
\label{fig4}
\end{figure}

Figure~\ref{fig4} shows that the EI phase boundary in the weak-coupling as well as in the strong-coupling regime is reproduced by poles of the uniform static excitonic susceptibility, as demonstrated analytically in Sec.~\ref{Gaussian-fluct}.
To determine the region where free excitons can exist, we evaluate the condition for exciton formation~\eqref{X-cond} subjected to the constraint~\eqref{X-cond_energy}. The exciton binding energy has to be positive. For numerical reasons we set the threshold to $\min(E_X^B)=10^{-6}$.  For the 3D case we restrict ourselves to excitons with a center-of-mass momentum ${\bf q}=0$.
In Fig.~\ref{fig4} the critical Coulomb interaction strength for the exciton formation $U_X(T)$ is shown by blue circles, where zero-momentum excitonic states exist for $U\geq U_X$. For low temperatures $U_X(T)$ coincides with the EI phase boundary in the strong-coupling regime. This suggests that in this regime the EI is developed by a BEC of preformed excitons. Within our analysis, for high temperatures $U_X(T)$ deviates slightly from the critical $U_g(T)$, determined from Eq.~\eqref{Hartree}, which separates the SM ($U\leq U_g$) and the SC ($U>U_g$), except for the point where $U_g(T)$ coincides with the EI phase boundary. This deviation turns out to be a result of the required finite binding energy of the excitons. In a SM excitons with ${\bf q}=0$ cannot exist. Here, the EI develops due to a BCS-type pairing of electrons and holes. We suggest that the BCS-BEC crossover region is estimated by extrapolating $U_g(T)$ into the EI phase.

To consider excitons with an arbitrary center-of-mass momentum, the ${\bf q}$-resolved excitonic susceptibility is calculated for the 2D EFKM, in order to keep the numerical effort manageable. Again we choose the band-structure parameters $E_\uparrow=-2.4$ and $t_\uparrow=-0.8$, for which the EI phase is stable in 2D,~\cite{BGBL04, Fa08} see Fig.~\ref{fig1}. To evaluate the SB parameters, the ${\bf k}$-summation is transformed into an energy integral using the tight-binding DOS for a square lattice. For the computation of the excitonic susceptibility the ${\bf k}$-summation is explicitly performed.

\begin{figure}[h]
\includegraphics[width=0.95\linewidth]{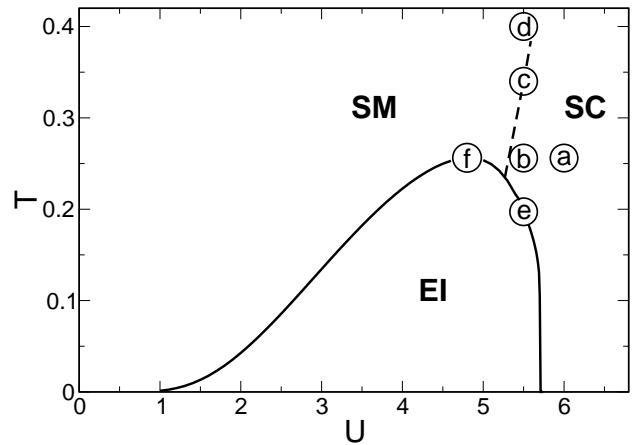}
\caption{Phase diagram of the 2D EFKM for the band-structure parameters $E_\uparrow=-2.4$ and $t_\uparrow=-0.8$. The solid line shows the EI phase boundary and the dashed line separates the SM and the SC phase. The exciton dispersion at the marked points a,b,c,d is shown in Fig.~\ref{fig8}, and the exciton dispersion for e and f is shown in Fig.~\ref{fig9}.}
\label{fig5}
\end{figure}
The phase diagram is shown in Fig~\ref{fig5}. Qualitatively there is no difference between the phase diagram of the 2D and 3D EFKM. Hence, the following results for 2D should hold qualitatively for 3D, too.

\begin{figure}[h]
\includegraphics[width=0.95\linewidth]{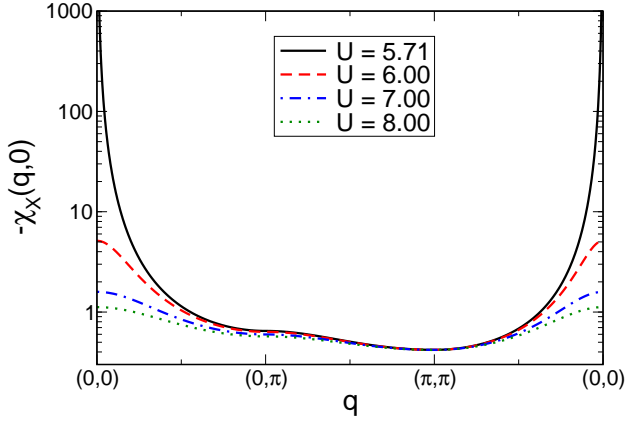}
\caption{(Color online) Static excitonic susceptibility for $T=0$ as a function of ${\bf q}$ (2D). For all $U$ we obtain $n_\uparrow=1$ and $n_\downarrow=0$.}
\label{fig6}
\end{figure}

Figure~\ref{fig6} shows the static excitonic susceptibility for zero temperature.
The formation of the EI phase is indicated by the divergence of $\chi_X({\bf q},0)$ at ${\bf q}=0$ for the critical value $U_{\rm EI}=5.71$. For $U>U_{\rm EI}$  the static excitonic susceptibility remains finite for all ${\bf q}$.

\subsection{Excitonic bound states}\label{excitons}
Next we want to characterize the paraphase in the vicinity of the EI phase with a view to the formation of uncondensed excitons.
\begin{figure}[h]
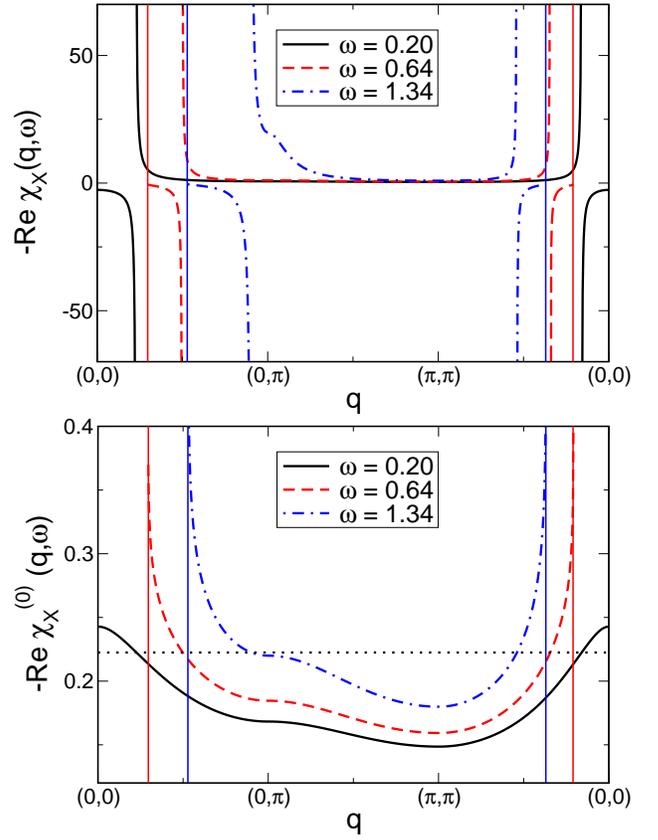

\includegraphics[width=0.95\linewidth]{fig7a.eps}
\includegraphics[width=0.95\linewidth]{fig7b.eps}
\caption{(Color online) Real part of the dynamic excitonic susceptibility as a function of ${\bf q}$ (2D) for $U=5.5$ and $T=0.256$ (upper panel) and the corresponding $\Re\chi_X^{(0)}({\bf q},\omega)$ (lower panel). The densities are $n_\uparrow=0.994$ and $n_\downarrow=0.006$. The vertical solid lines show the lower boundary of the electron-hole excitation continuum. The black dotted line gives the critical value of $\Re\chi_X^{(0)}$ for the exciton formation.}
\label{fig7}
\end{figure}
Figure~\ref{fig7} shows the real part of the dynamic excitonic susceptibility outside the continuum for several values of $\omega$. 
The divergences of $\Re\chi_X({\bf q},\omega)$ point out that the system is unstable against the formation of free excitons. With increasing excitation energy $\omega$ the exciton momentum ${\bf q}$ for the exciton formation increases due to the direct band gap situation. The divergence of $\Re\chi_X^{(0)}({\bf q},\omega)$ means $\omega=\omega_C({\bf q})$, shown as the vertical solid lines in Fig.~\ref{fig7}. For the ${\bf q}$ values where $\Re\chi_X({\bf q},\omega)$ (upper panel) and  $\Re\chi_X^{(0)}({\bf q},\omega)$ (lower panel) is not plotted in Fig.~\ref{fig7}, the given $\omega$ is larger than $\omega_C({\bf q})$. Hence, these ${\bf q}$ values are irrelevant for the exciton formation for the considered excitation energy $\omega$.

\begin{figure}[h]
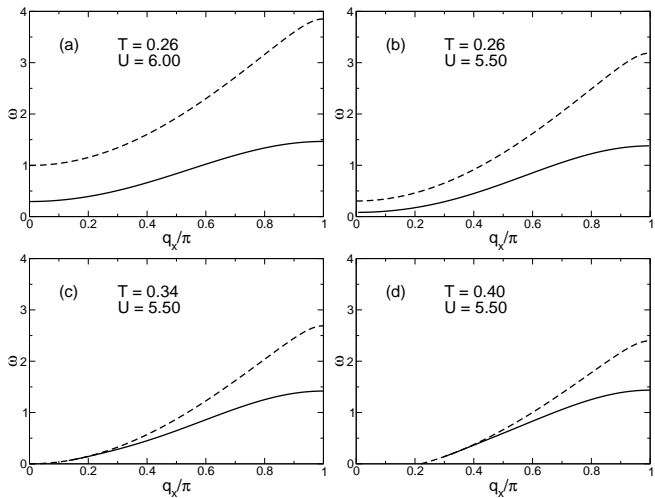

\begin{minipage}{0.49\linewidth}
\includegraphics[width=\linewidth]{fig8a.eps}
\end{minipage}
\begin{minipage}{0.49\linewidth}
\includegraphics[width=\linewidth]{fig8b.eps}
\end{minipage}
\begin{minipage}{0.49\linewidth}
\includegraphics[width=\linewidth]{fig8c.eps}
\end{minipage}
\begin{minipage}{0.49\linewidth}
\includegraphics[width=\linewidth]{fig8d.eps}
\end{minipage}
\caption{Exciton band (solid line) in comparison with the continuum band (dashed line) at the points marked in Fig.~\ref{fig5}: (a) in the SC phase  ($n_\uparrow=0.996$, $n_\downarrow=0.004$, $\tilde{\mu}=0.527$), (b) in the SC phase with a smaller band gap ($n_\uparrow=0.988$, $n_\downarrow=0.012$, $\tilde{\mu}=0.188$), (c) at the SC-SM transition ($n_\uparrow=0.973$, $n_\downarrow=0.027$, $\tilde{\mu}=0.056$), (d) in the SM phase ($n_\uparrow=0.960$, $n_\downarrow=0.040$, $\tilde{\mu}=-0.013$). The chemical potentials are measured relative to the valence band top. In all figures $q_y=0$ (2D).}
\label{fig8}
\end{figure}

Having confirmed the existence of excitonic states, we now turn to the properties of these states.
We denote the energy-momentum relation derived from Eq.~\eqref{X-cond} as the dispersion of the exciton band.
The smallest pole of $\Re\chi_X^{(0)}({\bf q},\omega)$ defines the ${\bf q}$-resolved lower boundary of the electron-hole excitation continuum, which we denote in the following as the continuum band.
In Fig.~\ref{fig8} the exciton band $\omega_X({\bf q})$ and the continuum band $\omega_C({\bf q})$ are shown for characteristic points in the phase diagram (see Fig.~\ref{fig5}).
In general, the continuum band is more sensitive to the temperature and Coulomb strength than the exciton band. Hence, the binding energy of the excitons is mainly affected by the continuum band when $T$ or $U$ is varied.  Figure~\ref{fig8} suggests that the exciton dispersion has a cosine-like form, determined by the underlying lattice.

In Fig.~\ref{fig8}(a), for any momentum a finite energy is needed to transfer one electron from the valence band to the conduction band, $\omega_C({\bf q})>0$, which is characteristic for the underlying SC band structure. Likewise the creation of an exciton requires energy, $\omega_X({\bf q})>0$ for all ${\bf q}$.
The comparison of Fig.~\ref{fig8}(b) with Fig.~\ref{fig8}(a) unveils the influence of the Coulomb interaction strength. In Fig.~\ref{fig8}(b), the exciton band is shifted to lower energies because the point (b) in the phase diagram (see Fig.~\ref{fig5}) is closer to the EI phase than (a). For the continuum bands $\omega_C^{(b)}({\bf q})<\omega_C	^{(a)}({\bf q})$ and, therefore, the binding energy of the excitons in Fig.~\ref{fig8}(b) is smaller than in Fig.~\ref{fig8}(a), since $U_{(b)}<U_{(a)}$, i.e., the Coulomb attraction between electrons and holes in Fig.~\ref{fig8}(b) is smaller than in Fig.~\ref{fig8}(a) as well, and the electrons and holes are more loosely bound.

The exciton and continuum dispersion at the SC-SM transition are shown in Fig.~\ref{fig8}(c). The continuum band reaches $\omega=0$ for ${\bf q}=0$, due to the direct band gap situation. The excitonic band disappears for small, finite center-of-mass momenta, not only for ${\bf q}=0$. The reason is the required finite binding energy of the excitons, i.e., $\omega_X({\bf q})$ should be separated from $\omega_C({\bf q})$.

In Fig.~\ref{fig8}(d) the SM band structure is characterized by the vanishing of positive $\omega_C({\bf q})$ at some finite value of ${\bf q}$. In a SM band structure excitonic states exist only with finite center-of-mass momenta, in contrast to Figs.~\ref{fig8}(a) and~\ref{fig8}(b). The existence of excitonic bound states in a SM is confirmed by  a very recent EFKM study with the projector-based renormalization method.~\cite{PBF11} The comparison of Figs.~\ref{fig8}(b), \ref{fig8}(c), and \ref{fig8}(d) demonstrates that the energetic position of the continuum band decreases with increasing temperature and, as a result, the exciton binding energy is lowered.

The qualitatively different feature of the exciton and the continuum band in the SC and in the SM phase suggests that the condensation process at the SC-EI transition differs from the one at the SM-EI transition.
\begin{figure}[h]
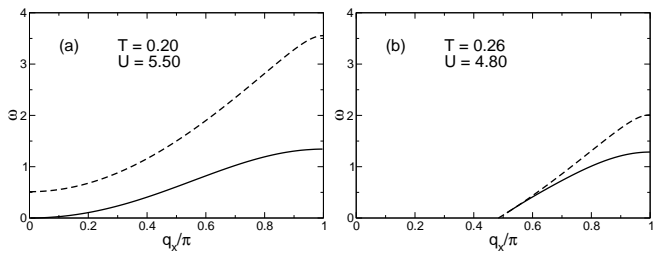

\begin{minipage}{0.49\linewidth}
\includegraphics[width=\linewidth]{fig9a.eps}
\end{minipage}
\begin{minipage}{0.49\linewidth}
\includegraphics[width=\linewidth]{fig9b.eps}
\end{minipage}
\caption{Exciton band (solid line) and continuum band (dashed line) (a) at point {\large \textcircled{\normalsize e}} marked in Fig.~\ref{fig5}:  the SC-EI transition ($n_\uparrow=0.995$, $n_\downarrow=0.005$, $\tilde{\mu}=0.280$), and (b) at point{\large \textcircled{\normalsize f}} marked in Fig.~\ref{fig5}: the SM-EI transition  ($n_\uparrow=0.949$, $n_\downarrow=0.051$, $\tilde{\mu}=-0.375$). The chemical potentials are measured relative to the valence band top. In all figures $q_y=0$.}
\label{fig9}
\end{figure}

Figure~\ref{fig9}(a) shows the exciton and the continuum dispersion at the SC-EI transition.
As already stated, $\omega_C({\bf q})>0$ $\forall {\bf q}$ reflects the underlying SC band structure. The minimum of the exciton band (at ${\bf q}=0$) reaches $\omega=0$, so that the creation of a zero-momentum exciton does not require energy. Physically, comparing only the energies the exciton band touches the valence band top. In our analysis, this is the point where the BEC of excitons sets in. 

On the contrary, the exciton dispersion at the SM-EI transition, shown in Fig.~\ref{fig9}(b), gives no hint for a condensation process. In fact, it is a characteristic plot for a SM band structure: there are excitonic bound states, but only with a finite center-of-mass momentum, c.f. Fig.~\ref{fig8}(d). In the SM regime there is a large and well-defined Fermi surface and the electron-hole condensation at the SM-EI transition can be imagined as a BCS-type process.

\section{Summary}\label{summary}
In this paper we studied the extended Falicov-Kimball model within a $SO(2)$-invariant slave-boson treatment taking Gaussian fluctuations into account. We computed the static charge and orbital susceptibility in order to investigate the formation of a charge density wave and staggered orbital order. Analyzing the static excitonic susceptibility, the instability towards an excitonic insulator (EI) phase agrees with the saddle-point phase diagram. 
By calculating the dynamic excitonic susceptibility, we analyzed the formation of excitons in the phase without long-range order. We found that finite-momentum excitons form in the vicinity of the EI phase, not only at the semiconducting (SC) side but also at the semimetal (SM) side. At the transition to the EI phase we observed qualitatively different features at the SC and the SM side. At the SC-EI transition the excitation energy of the excitonic state with zero center-of-mass momentum vanishes, leading to a Bose-Einstein condensation of these preformed excitons. At the SM side there are no excitonic bound states with ${\bf q}=0$.  Here, the EI forms due to a BCS-type pairing of electrons and holes, and the occurring excitonic states of finite momentum are not of importance for the phase transition.

\section*{Acknowledgments}
The authors thank K. W. Becker,  R. L. Heinisch, V.-N. Phan, and O. P. Shuskov for stimulating 
discussions. HF acknowledges a Gordon Godfrey professorial 
visiting fellowship at the UNSW, Australia.
This work was supported by DFG through SFB 652. 


\begin{thebibliography}{38}%
\makeatletter
\providecommand \@ifxundefined [1]{%
 \@ifx{#1\undefined}
}%
\providecommand \@ifnum [1]{%
 \ifnum #1\expandafter \@firstoftwo
 \else \expandafter \@secondoftwo
 \fi
}%
\providecommand \@ifx [1]{%
 \ifx #1\expandafter \@firstoftwo
 \else \expandafter \@secondoftwo
 \fi
}%
\providecommand \natexlab [1]{#1}%
\providecommand \enquote  [1]{``#1''}%
\providecommand \bibnamefont  [1]{#1}%
\providecommand \bibfnamefont [1]{#1}%
\providecommand \citenamefont [1]{#1}%
\providecommand \href@noop [0]{\@secondoftwo}%
\providecommand \href [0]{\begingroup \@sanitize@url \@href}%
\providecommand \@href[1]{\@@startlink{#1}\@@href}%
\providecommand \@@href[1]{\endgroup#1\@@endlink}%
\providecommand \@sanitize@url [0]{\catcode `\\12\catcode `\$12\catcode
  `\&12\catcode `\#12\catcode `\^12\catcode `\_12\catcode `\%12\relax}%
\providecommand \@@startlink[1]{}%
\providecommand \@@endlink[0]{}%
\providecommand \url  [0]{\begingroup\@sanitize@url \@url }%
\providecommand \@url [1]{\endgroup\@href {#1}{\urlprefix }}%
\providecommand \urlprefix  [0]{URL }%
\providecommand \Eprint [0]{\href }%
\providecommand \doibase [0]{http://dx.doi.org/}%
\providecommand \selectlanguage [0]{\@gobble}%
\providecommand \bibinfo  [0]{\@secondoftwo}%
\providecommand \bibfield  [0]{\@secondoftwo}%
\providecommand \translation [1]{[#1]}%
\providecommand \BibitemOpen [0]{}%
\providecommand \bibitemStop [0]{}%
\providecommand \bibitemNoStop [0]{.\EOS\space}%
\providecommand \EOS [0]{\spacefactor3000\relax}%
\providecommand \BibitemShut  [1]{\csname bibitem#1\endcsname}%
\let\auto@bib@innerbib\@empty
\bibitem [{\citenamefont {Halperin}\ and\ \citenamefont {Rice}(1968)}]{HR68}%
  \BibitemOpen
  \bibfield  {author} {\bibinfo {author} {\bibfnamefont {B.~I.}\ \bibnamefont
  {Halperin}}\ and\ \bibinfo {author} {\bibfnamefont {T.~M.}\ \bibnamefont
  {Rice}},\ }\href@noop {} {\bibfield  {journal} {\bibinfo  {journal} {Rev.
  Mod. Phys.}\ }\textbf {\bibinfo {volume} {40}},\ \bibinfo {pages} {755}
  (\bibinfo {year} {1968})}\BibitemShut {NoStop}%
%
%
\bibitem [{\citenamefont {Mott}(1961)}]{M61}%
  \BibitemOpen
  \bibfield  {author} {\bibinfo {author} {\bibfnamefont {N.~F.}\ \bibnamefont
  {Mott}},\ }\href@noop {} {\bibfield  {journal} {\bibinfo  {journal} {Philos.
  Mag.}\ }\textbf {\bibinfo {volume} {6}},\ \bibinfo {pages} {287} (\bibinfo
  {year} {1961})}\BibitemShut {NoStop}%
%
%
\bibitem [{\citenamefont {Knox}(1963)}]{Kno63}%
  \BibitemOpen
  \bibfield  {author} {\bibinfo {author} {\bibfnamefont {R.}~\bibnamefont
  {Knox}},\ }in\ \href@noop {} {\emph {\bibinfo {booktitle} {Solid State
  Physics}}},\ \bibinfo {editor} {edited by\ \bibinfo {editor} {\bibfnamefont
  {F.}~\bibnamefont {Seitz}}\ and\ \bibinfo {editor} {\bibfnamefont
  {D.}~\bibnamefont {Turnbull}}}\ (\bibinfo  {publisher} {Academic Press},\
  \bibinfo {address} {New York},\ \bibinfo {year} {1963})\ p.\ \bibinfo {pages}
  {Suppl. 5 p. 100}\BibitemShut {NoStop}%
%
%
\bibitem [{\citenamefont {Comte}\ and\ \citenamefont
  {Nozi\`{e}res}(1982)}]{CN82a}%
  \BibitemOpen
  \bibfield  {author} {\bibinfo {author} {\bibfnamefont {C.}~\bibnamefont
  {Comte}}\ and\ \bibinfo {author} {\bibfnamefont {P.}~\bibnamefont
  {Nozi\`{e}res}},\ }\href@noop {} {\bibfield  {journal} {\bibinfo  {journal}
  {J. Phys. (France)}\ }\textbf {\bibinfo {volume} {43}},\ \bibinfo {pages}
  {1069} (\bibinfo {year} {1982})}\BibitemShut {NoStop}%
%
%
\bibitem [{\citenamefont {Bronold}\ and\ \citenamefont {Fehske}(2006)}]{BF06}%
  \BibitemOpen
  \bibfield  {author} {\bibinfo {author} {\bibfnamefont {F.~X.}\ \bibnamefont
  {Bronold}}\ and\ \bibinfo {author} {\bibfnamefont {H.}~\bibnamefont
  {Fehske}},\ }\href@noop {} {\bibfield  {journal} {\bibinfo  {journal} {Phys.
  Rev. B}\ }\textbf {\bibinfo {volume} {74}},\ \bibinfo {pages} {165107}
  (\bibinfo {year} {2006})}\BibitemShut {NoStop}%
%
%
\bibitem [{\citenamefont {Ihle}\ \emph {et~al.}(2008)\citenamefont {Ihle},
  \citenamefont {Pfafferott}, \citenamefont {Burovski}, \citenamefont
  {Bronold},\ and\ \citenamefont {Fehske}}]{IPBBF08}%
  \BibitemOpen
  \bibfield  {author} {\bibinfo {author} {\bibfnamefont {D.}~\bibnamefont
  {Ihle}}, \bibinfo {author} {\bibfnamefont {M.}~\bibnamefont {Pfafferott}},
  \bibinfo {author} {\bibfnamefont {E.}~\bibnamefont {Burovski}}, \bibinfo
  {author} {\bibfnamefont {F.~X.}\ \bibnamefont {Bronold}}, \ and\ \bibinfo
  {author} {\bibfnamefont {H.}~\bibnamefont {Fehske}},\ }\href@noop {}
  {\bibfield  {journal} {\bibinfo  {journal} {Phys. Rev. B}\ }\textbf {\bibinfo
  {volume} {78}},\ \bibinfo {pages} {193103} (\bibinfo {year}
  {2008})}\BibitemShut {NoStop}%
%
%
\bibitem [{\citenamefont {Phan}\ \emph {et~al.}(2010)\citenamefont {Phan},
  \citenamefont {Becker},\ and\ \citenamefont {Fehske}}]{PBF10}%
  \BibitemOpen
  \bibfield  {author} {\bibinfo {author} {\bibfnamefont {V.-N.}\ \bibnamefont
  {Phan}}, \bibinfo {author} {\bibfnamefont {K.~W.}\ \bibnamefont {Becker}}, \
  and\ \bibinfo {author} {\bibfnamefont {H.}~\bibnamefont {Fehske}},\
  }\href@noop {} {\bibfield  {journal} {\bibinfo  {journal} {Phys. Rev. B}\
  }\textbf {\bibinfo {volume} {81}},\ \bibinfo {pages} {205117} (\bibinfo
  {year} {2010})}\BibitemShut {NoStop}%
%
%
\bibitem [{\citenamefont {{des Cloizeaux}}(1965)}]{C65}%
  \BibitemOpen
  \bibfield  {author} {\bibinfo {author} {\bibfnamefont {J.}~\bibnamefont {{des
  Cloizeaux}}},\ }\href@noop {} {\bibfield  {journal} {\bibinfo  {journal} {J.
  Phys. Chem. Solids}\ }\textbf {\bibinfo {volume} {26}},\ \bibinfo {pages}
  {259} (\bibinfo {year} {1965})}; 
  {\bibfield  {author} {\bibinfo {author} {\bibfnamefont {L.~V.}\ \bibnamefont
  {Keldysh}}\ and\ \bibinfo {author} {\bibfnamefont {H.~Y.~V.}\ \bibnamefont
  {Kopaev}},\ }}\href@noop {} {\bibfield  {journal} {\bibinfo  {journal} {Sov.
  Phys. Sol. State}\ }\textbf {\bibinfo {volume} {6}},\ \bibinfo {pages} {2219}
  (\bibinfo {year} {1965})}; 
  {\bibfield  {author} {\bibinfo {author} {\bibfnamefont {D.}~\bibnamefont
  {J\'{e}rome}}, \bibinfo {author} {\bibfnamefont {T.~M.}\ \bibnamefont
  {Rice}}, \ and\ \bibinfo {author} {\bibfnamefont {W.}~\bibnamefont {Kohn}}},\
  }\href@noop {} {\bibfield  {journal} {\bibinfo  {journal} {Physical Review}\
  }\textbf {\bibinfo {volume} {158}},\ \bibinfo {pages} {462} (\bibinfo {year}
  {1967})}\BibitemShut {NoStop}%
%
%
\bibitem [{\citenamefont {Littlewood}\ \emph {et~al.}(2004)\citenamefont
  {Littlewood}, \citenamefont {Eastham}, \citenamefont {Keeling}, \citenamefont
  {Marchetti}, \citenamefont {Simons},\ and\ \citenamefont
  {Szymanska}}]{LEKMSS04}%
  \BibitemOpen
  \bibfield  {author} {\bibinfo {author} {\bibfnamefont {P.~B.}\ \bibnamefont
  {Littlewood}}, \bibinfo {author} {\bibfnamefont {P.~R.}\ \bibnamefont
  {Eastham}}, \bibinfo {author} {\bibfnamefont {J.~M.~J.}\ \bibnamefont
  {Keeling}}, \bibinfo {author} {\bibfnamefont {F.~M.}\ \bibnamefont
  {Marchetti}}, \bibinfo {author} {\bibfnamefont {B.~D.}\ \bibnamefont
  {Simons}}, \ and\ \bibinfo {author} {\bibfnamefont {M.~H.}\ \bibnamefont
  {Szymanska}},\ }\href@noop {} {\bibfield  {journal} {\bibinfo  {journal} {J.
  Phys. Condens. Matter}\ }\textbf {\bibinfo {volume} {16}} (\bibinfo {year}
  {2004})}; 
  {\bibfield  {author} {\bibinfo {author} {\bibfnamefont {C.}~\bibnamefont
  {Monney}}, \bibinfo {author} {\bibfnamefont {E.~F.}\ \bibnamefont {Schwier}},
  \bibinfo {author} {\bibfnamefont {M.~G.}\ \bibnamefont {Garnier}}, \bibinfo
  {author} {\bibfnamefont {N.}~\bibnamefont {Mariotti}}, \bibinfo {author}
  {\bibfnamefont {C.}~\bibnamefont {Didiot}}, \bibinfo {author} {\bibfnamefont
  {H.}~\bibnamefont {Cercellier}}, \bibinfo {author} {\bibfnamefont
  {J.}~\bibnamefont {Marcus}}, \bibinfo {author} {\bibfnamefont
  {H.}~\bibnamefont {Berger}}, \bibinfo {author} {\bibfnamefont {A.~N.}\
  \bibnamefont {Titov}}, \bibinfo {author} {\bibfnamefont {H.}~\bibnamefont
  {Beck}}, \ and\ \bibinfo {author} {\bibfnamefont {P.}~\bibnamefont {Aebi}}},\
  }\href@noop {} {\bibfield  {journal} {\bibinfo  {journal} {New J. Phys.}\
  }\textbf {\bibinfo {volume} {12}},\ \bibinfo {pages} {125019} (\bibinfo
  {year} {2010}{\natexlab{a}})}\BibitemShut {NoStop}%
%
%
\bibitem [{\citenamefont {Neuenschwander}\ and\ \citenamefont
  {Wachter}(1990)}]{NW90}%
  \BibitemOpen
  \bibfield  {author} {\bibinfo {author} {\bibfnamefont {J.}~\bibnamefont
  {Neuenschwander}}\ and\ \bibinfo {author} {\bibfnamefont {P.}~\bibnamefont
  {Wachter}},\ }\href@noop {} {\bibfield  {journal} {\bibinfo  {journal} {Phys.
  Rev. B}\ }\textbf {\bibinfo {volume} {41}},\ \bibinfo {pages} {12693}
  (\bibinfo {year} {1990})}; 
  {\bibfield  {author} {\bibinfo {author} {\bibfnamefont {B.}~\bibnamefont
  {Bucher}}, \bibinfo {author} {\bibfnamefont {P.}~\bibnamefont {Steiner}}, \
  and\ \bibinfo {author} {\bibfnamefont {P.}~\bibnamefont {Wachter}}},\
  }\href@noop {} {\bibfield  {journal} {\bibinfo  {journal} {Phys. Rev. Lett.}\
  }\textbf {\bibinfo {volume} {67}},\ \bibinfo {pages} {2717} (\bibinfo {year}
  {1991})}; 
  {\bibfield  {author} {\bibinfo {author} {\bibfnamefont {P.}~\bibnamefont
  {Wachter}}, \bibinfo {author} {\bibfnamefont {B.}~\bibnamefont {Bucher}}, \
  and\ \bibinfo {author} {\bibfnamefont {J.}~\bibnamefont {Malar}}},\
  }\href@noop {} {\bibfield  {journal} {\bibinfo  {journal} {Phys. Rev. B}\
  }\textbf {\bibinfo {volume} {69}},\ \bibinfo {pages} {094502} (\bibinfo
  {year} {2004})}\BibitemShut {NoStop}%
%
%
\bibitem [{\citenamefont {Wakisaka}\ \emph {et~al.}(2009)\citenamefont
  {Wakisaka}, \citenamefont {Sudayama}, \citenamefont {Takubo}, \citenamefont
  {Mizokawa}, \citenamefont {Arita}, \citenamefont {Namatame}, \citenamefont
  {Taniguchi}, \citenamefont {Katayama}, \citenamefont {Nohara},\ and\
  \citenamefont {Takagi}}]{WaEtAl09}%
  \BibitemOpen
  \bibfield  {author} {\bibinfo {author} {\bibfnamefont {Y.}~\bibnamefont
  {Wakisaka}}, \bibinfo {author} {\bibfnamefont {T.}~\bibnamefont {Sudayama}},
  \bibinfo {author} {\bibfnamefont {K.}~\bibnamefont {Takubo}}, \bibinfo
  {author} {\bibfnamefont {T.}~\bibnamefont {Mizokawa}}, \bibinfo {author}
  {\bibfnamefont {M.}~\bibnamefont {Arita}}, \bibinfo {author} {\bibfnamefont
  {H.}~\bibnamefont {Namatame}}, \bibinfo {author} {\bibfnamefont
  {M.}~\bibnamefont {Taniguchi}}, \bibinfo {author} {\bibfnamefont
  {N.}~\bibnamefont {Katayama}}, \bibinfo {author} {\bibfnamefont
  {M.}~\bibnamefont {Nohara}}, \ and\ \bibinfo {author} {\bibfnamefont
  {H.}~\bibnamefont {Takagi}},\ }\href@noop {} {\bibfield  {journal} {\bibinfo
  {journal} {Phys. Rev. Lett.}\ }\textbf {\bibinfo {volume} {103}},\ \bibinfo
  {pages} {026402} (\bibinfo {year} {2009})}\BibitemShut {NoStop}%
%
%
\bibitem [{\citenamefont {Cercellier}\ \emph {et~al.}(2007)\citenamefont
  {Cercellier}, \citenamefont {Monney}, \citenamefont {Clerc}, \citenamefont
  {Battaglia}, \citenamefont {Garnier}, \citenamefont {Beck},\ and\
  \citenamefont {Aebi}}]{CeEtAl07}%
  \BibitemOpen
  \bibfield  {author} {\bibinfo {author} {\bibfnamefont {H.}~\bibnamefont
  {Cercellier}}, \bibinfo {author} {\bibfnamefont {C.}~\bibnamefont {Monney}},
  \bibinfo {author} {\bibfnamefont {F.}~\bibnamefont {Clerc}}, \bibinfo
  {author} {\bibfnamefont {C.}~\bibnamefont {Battaglia}}, \bibinfo {author}
  {\bibfnamefont {M.~G.}\ \bibnamefont {Garnier}}, \bibinfo {author}
  {\bibfnamefont {H.}~\bibnamefont {Beck}}, \ and\ \bibinfo {author}
  {\bibfnamefont {P.}~\bibnamefont {Aebi}},\ }\href@noop {} {\bibfield
  {journal} {\bibinfo  {journal} {Phys. Rev. Lett.}\ }\textbf {\bibinfo
  {volume} {99}},\ \bibinfo {pages} {146403} (\bibinfo {year}
  {2007})}; 
  {\bibfield  {author} {\bibinfo {author} {\bibfnamefont {C.}~\bibnamefont
  {Monney}}, \bibinfo {author} {\bibfnamefont {E.~F.}\ \bibnamefont {Schwier}},
  \bibinfo {author} {\bibfnamefont {C.}~\bibnamefont {Battaglia}}, \bibinfo
  {author} {\bibfnamefont {M.~G.}\ \bibnamefont {Garnier}}, \bibinfo {author}
  {\bibfnamefont {N.}~\bibnamefont {Mariotti}}, \bibinfo {author}
  {\bibfnamefont {C.}~\bibnamefont {Didiot}}, \bibinfo {author} {\bibfnamefont
  {H.}~\bibnamefont {Beck}}, \bibinfo {author} {\bibfnamefont {P.}~\bibnamefont
  {Aebi}}, \bibinfo {author} {\bibfnamefont {H.}~\bibnamefont {Cercellier}},
  \bibinfo {author} {\bibfnamefont {J.}~\bibnamefont {Marcus}}, \bibinfo
  {author} {\bibfnamefont {H.}~\bibnamefont {Berger}}, \ and\ \bibinfo {author}
  {\bibfnamefont {A.~N.}\ \bibnamefont {Titov}}},\ }\href@noop {} {\bibfield
  {journal} {\bibinfo  {journal} {Phys. Rev. B}\ }\textbf {\bibinfo {volume}
  {81}},\ \bibinfo {pages} {155104} (\bibinfo {year}
  {2010}{\natexlab{b}})}\BibitemShut {NoStop}%
%
%
\bibitem [{\citenamefont {Falicov}\ and\ \citenamefont {Kimball}(1969)}]{FK69}%
  \BibitemOpen
  \bibfield  {author} {\bibinfo {author} {\bibfnamefont {L.~M.}\ \bibnamefont
  {Falicov}}\ and\ \bibinfo {author} {\bibfnamefont {J.~C.}\ \bibnamefont
  {Kimball}},\ }\href@noop {} {\bibfield  {journal} {\bibinfo  {journal} {Phys.
  Rev. Lett.}\ }\textbf {\bibinfo {volume} {22}},\ \bibinfo {pages} {997}
  (\bibinfo {year} {1969})}; 
  {\bibfield  {author} {\bibinfo {author} {\bibfnamefont {R.}~\bibnamefont
  {Ramirez}}, \bibinfo {author} {\bibfnamefont {L.~M.}\ \bibnamefont
  {Falicov}}, \ and\ \bibinfo {author} {\bibfnamefont {J.~C.}\ \bibnamefont
  {Kimball}}},\ }\href@noop {} {\bibfield  {journal} {\bibinfo  {journal} {Phys.
  Rev. B}\ }\textbf {\bibinfo {volume} {2}},\ \bibinfo {pages} {3383} (\bibinfo
  {year} {1970})}\BibitemShut {NoStop}%
%
%
\bibitem [{\citenamefont {Subrahmanyam}\ and\ \citenamefont
  {Barma}(1988)}]{SB88}%
  \BibitemOpen
  \bibfield  {author} {\bibinfo {author} {\bibfnamefont {V.}~\bibnamefont
  {Subrahmanyam}}\ and\ \bibinfo {author} {\bibfnamefont {M.}~\bibnamefont
  {Barma}},\ }\href@noop {} {\bibfield  {journal} {\bibinfo  {journal} {J.
  Phys. Chem.}\ }\textbf {\bibinfo {volume} {21}},\ \bibinfo {pages} {L19}
  (\bibinfo {year} {1988})}\BibitemShut {NoStop}%
%
%
\bibitem [{\citenamefont {Kanda}\ \emph {et~al.}(1976)\citenamefont {Kanda},
  \citenamefont {Machida},\ and\ \citenamefont {Matsubara}}]{KMM76}%
  \BibitemOpen
  \bibfield  {author} {\bibinfo {author} {\bibfnamefont {K.}~\bibnamefont
  {Kanda}}, \bibinfo {author} {\bibfnamefont {K.}~\bibnamefont {Machida}}, \
  and\ \bibinfo {author} {\bibfnamefont {T.}~\bibnamefont {Matsubara}},\
  }\href@noop {} {\bibfield  {journal} {\bibinfo  {journal} {Solid State
  Commun.}\ }\textbf {\bibinfo {volume} {19}},\ \bibinfo {pages} {651}
  (\bibinfo {year} {1976})}\BibitemShut {NoStop}%
%
%
\bibitem [{\citenamefont {Batista}(2002)}]{Ba02b}%
  \BibitemOpen
  \bibfield  {author} {\bibinfo {author} {\bibfnamefont {C.~D.}\ \bibnamefont
  {Batista}},\ }\href@noop {} {\bibfield  {journal} {\bibinfo  {journal} {Phys.
  Rev. Lett.}\ }\textbf {\bibinfo {volume} {89}},\ \bibinfo {pages} {166403}
  (\bibinfo {year} {2002})}\BibitemShut {NoStop}%
%
%
\bibitem [{\citenamefont {Batista}\ \emph {et~al.}(2004)\citenamefont
  {Batista}, \citenamefont {Gubernatis}, \citenamefont {Bon\v{c}a},\ and\
  \citenamefont {Lin}}]{BGBL04}%
  \BibitemOpen
  \bibfield  {author} {\bibinfo {author} {\bibfnamefont {C.~D.}\ \bibnamefont
  {Batista}}, \bibinfo {author} {\bibfnamefont {J.~E.}\ \bibnamefont
  {Gubernatis}}, \bibinfo {author} {\bibfnamefont {J.}~\bibnamefont
  {Bon\v{c}a}}, \ and\ \bibinfo {author} {\bibfnamefont {H.~Q.}\ \bibnamefont
  {Lin}},\ }\href@noop {} {\bibfield  {journal} {\bibinfo  {journal} {Phys.
  Rev. Lett.}\ }\textbf {\bibinfo {volume} {92}},\ \bibinfo {pages} {187601}
  (\bibinfo {year} {2004})}\BibitemShut {NoStop}%
%
%
\bibitem [{\citenamefont {Brydon}(2008)}]{Br08}%
  \BibitemOpen
  \bibfield  {author} {\bibinfo {author} {\bibfnamefont {P.~M.~R.}\
  \bibnamefont {Brydon}},\ }\href@noop {} {\bibfield  {journal} {\bibinfo
  {journal} {Phys. Rev. B}\ }\textbf {\bibinfo {volume} {77}},\ \bibinfo
  {pages} {045109} (\bibinfo {year} {2008})}\BibitemShut {NoStop}%
%
%
\bibitem [{\citenamefont {Zenker}\ \emph
  {et~al.}(2010{\natexlab{a}})\citenamefont {Zenker}, \citenamefont {Ihle},
  \citenamefont {Bronold},\ and\ \citenamefont {Fehske}}]{ZIBF10}%
  \BibitemOpen
  \bibfield  {author} {\bibinfo {author} {\bibfnamefont {B.}~\bibnamefont
  {Zenker}}, \bibinfo {author} {\bibfnamefont {D.}~\bibnamefont {Ihle}},
  \bibinfo {author} {\bibfnamefont {F.~X.}\ \bibnamefont {Bronold}}, \ and\
  \bibinfo {author} {\bibfnamefont {H.}~\bibnamefont {Fehske}},\ }\href@noop {}
  {\bibfield  {journal} {\bibinfo  {journal} {Phys. Rev. B}\ }\textbf {\bibinfo
  {volume} {81}},\ \bibinfo {pages} {115122} (\bibinfo {year}
  {2010}{\natexlab{a}})}\BibitemShut {NoStop}%
%
%
\bibitem [{\citenamefont {Farka\v{s}ovsk\'{y}}(2008)}]{Fa08}%
  \BibitemOpen
  \bibfield  {author} {\bibinfo {author} {\bibfnamefont {P.}~\bibnamefont
  {Farka\v{s}ovsk\'{y}}},\ }\href@noop {} {\bibfield  {journal} {\bibinfo
  {journal} {Phys. Rev. B}\ }\textbf {\bibinfo {volume} {77}},\ \bibinfo
  {pages} {155130} (\bibinfo {year} {2008})}\BibitemShut {NoStop}%
%
%
\bibitem [{\citenamefont {Schneider}\ and\ \citenamefont
  {Czycholl}(2008)}]{SC08}%
  \BibitemOpen
  \bibfield  {author} {\bibinfo {author} {\bibfnamefont {C.}~\bibnamefont
  {Schneider}}\ and\ \bibinfo {author} {\bibfnamefont {G.}~\bibnamefont
  {Czycholl}},\ }\href@noop {} {\bibfield  {journal} {\bibinfo  {journal} {Eur.
  Phys. J. B}\ }\textbf {\bibinfo {volume} {64}},\ \bibinfo {pages} {43}
  (\bibinfo {year} {2008})}\BibitemShut {NoStop}%
%
%
\bibitem [{\citenamefont {Gutzwiller}(1963)}]{Gu63}%
  \BibitemOpen
  \bibfield  {author} {\bibinfo {author} {\bibfnamefont {M.~C.}\ \bibnamefont
  {Gutzwiller}},\ }\href@noop {} {\bibfield  {journal} {\bibinfo  {journal}
  {Phys. Rev. Lett.}\ }\textbf {\bibinfo {volume} {10}},\ \bibinfo {pages}
  {159} (\bibinfo {year} {1963})}\BibitemShut {NoStop}%
%
%
\bibitem [{\citenamefont {Kotliar}\ and\ \citenamefont
  {Ruckenstein}(1986)}]{KR86}%
  \BibitemOpen
  \bibfield  {author} {\bibinfo {author} {\bibfnamefont {G.}~\bibnamefont
  {Kotliar}}\ and\ \bibinfo {author} {\bibfnamefont {A.~E.}\ \bibnamefont
  {Ruckenstein}},\ }\href@noop {} {\bibfield  {journal} {\bibinfo  {journal}
  {Phys. Rev. Lett.}\ }\textbf {\bibinfo {volume} {57}},\ \bibinfo {pages}
  {1362} (\bibinfo {year} {1986})}\BibitemShut {NoStop}%
%
%
\bibitem [{\citenamefont {Li}\ \emph {et~al.}(1989)\citenamefont {Li},
  \citenamefont {W\"olfle},\ and\ \citenamefont {Hirschfeld}}]{LWH89}%
  \BibitemOpen
  \bibfield  {author} {\bibinfo {author} {\bibfnamefont {T.}~\bibnamefont
  {Li}}, \bibinfo {author} {\bibfnamefont {P.}~\bibnamefont {W\"olfle}}, \ and\
  \bibinfo {author} {\bibfnamefont {P.~J.}\ \bibnamefont {Hirschfeld}},\
  }\href@noop {} {\bibfield  {journal} {\bibinfo  {journal} {Phys. Rev. B}\
  }\textbf {\bibinfo {volume} {40}},\ \bibinfo {pages} {6817} (\bibinfo {year}
  {1989})}\BibitemShut {NoStop}%
%
%
\bibitem [{\citenamefont {Lechermann}\ \emph {et~al.}(2007)\citenamefont
  {Lechermann}, \citenamefont {Georges}, \citenamefont {Kotliar},\ and\
  \citenamefont {Parcollet}}]{LGKP07}%
  \BibitemOpen
  \bibfield  {author} {\bibinfo {author} {\bibfnamefont {F.}~\bibnamefont
  {Lechermann}}, \bibinfo {author} {\bibfnamefont {A.}~\bibnamefont {Georges}},
  \bibinfo {author} {\bibfnamefont {G.}~\bibnamefont {Kotliar}}, \ and\
  \bibinfo {author} {\bibfnamefont {O.}~\bibnamefont {Parcollet}},\ }\href@noop
  {} {\bibfield  {journal} {\bibinfo  {journal} {Phys. Rev. B}\ }\textbf
  {\bibinfo {volume} {76}},\ \bibinfo {pages} {155102} (\bibinfo {year}
  {2007})}; 
  {\bibfield  {author} {\bibinfo {author} {\bibfnamefont {J.}~\bibnamefont
  {B\"{u}nemann}}},\ }\href@noop {} {\bibfield  {journal} {\bibinfo  {journal}
  {Phys. Status Solidi B}\ }\textbf {\bibinfo {volume} {248}},\ \bibinfo
  {pages} {203} (\bibinfo {year} {2011})}\BibitemShut {NoStop}%
%
%
\bibitem [{\citenamefont {Lavagna}(1990)}]{La90}%
  \BibitemOpen
  \bibfield  {author} {\bibinfo {author} {\bibfnamefont {M.}~\bibnamefont
  {Lavagna}},\ }\href@noop {} {\bibfield  {journal} {\bibinfo  {journal} {Phys.
  Rev. B}\ }\textbf {\bibinfo {volume} {41}},\ \bibinfo {pages} {142} (\bibinfo
  {year} {1990})}\BibitemShut {NoStop}%
%
%
\bibitem [{\citenamefont {Li}\ \emph {et~al.}(1991)\citenamefont {Li},
  \citenamefont {Sun},\ and\ \citenamefont {W\"olfle}}]{LSW91}%
  \BibitemOpen
  \bibfield  {author} {\bibinfo {author} {\bibfnamefont {T.}~\bibnamefont
  {Li}}, \bibinfo {author} {\bibfnamefont {Y.~S.}\ \bibnamefont {Sun}}, \ and\
  \bibinfo {author} {\bibfnamefont {P.}~\bibnamefont {W\"olfle}},\ }\href@noop
  {} {\bibfield  {journal} {\bibinfo  {journal} {Z. Phys. B}\ }\textbf
  {\bibinfo {volume} {82}},\ \bibinfo {pages} {369} (\bibinfo {year}
  {1991})}\BibitemShut {NoStop}%
%
%
\bibitem [{\citenamefont {Deeg}\ \emph {et~al.}(1994)\citenamefont {Deeg},
  \citenamefont {Fehske}, \citenamefont {K\"orner}, \citenamefont {Trimper},\
  and\ \citenamefont {Ihle}}]{DFKTI94}%
  \BibitemOpen
  \bibfield  {author} {\bibinfo {author} {\bibfnamefont {M.}~\bibnamefont
  {Deeg}}, \bibinfo {author} {\bibfnamefont {H.}~\bibnamefont {Fehske}},
  \bibinfo {author} {\bibfnamefont {S.}~\bibnamefont {K\"orner}}, \bibinfo
  {author} {\bibfnamefont {S.}~\bibnamefont {Trimper}}, \ and\ \bibinfo
  {author} {\bibfnamefont {D.}~\bibnamefont {Ihle}},\ }\href@noop {} {\bibfield
   {journal} {\bibinfo  {journal} {Z. Phys. B}\ }\textbf {\bibinfo {volume}
  {95}},\ \bibinfo {pages} {87} (\bibinfo {year} {1994})}\BibitemShut {NoStop}%
%
%
\bibitem [{\citenamefont {Zenker}\ \emph
  {et~al.}(2010{\natexlab{b}})\citenamefont {Zenker}, \citenamefont {Fehske},\
  and\ \citenamefont {Batista}}]{ZFB10}%
  \BibitemOpen
  \bibfield  {author} {\bibinfo {author} {\bibfnamefont {B.}~\bibnamefont
  {Zenker}}, \bibinfo {author} {\bibfnamefont {H.}~\bibnamefont {Fehske}}, \
  and\ \bibinfo {author} {\bibfnamefont {C.~D.}\ \bibnamefont {Batista}},\
  }\href@noop {} {\bibfield  {journal} {\bibinfo  {journal} {Phys. Rev. B}\
  }\textbf {\bibinfo {volume} {82}},\ \bibinfo {pages} {165110} (\bibinfo
  {year} {2010}{\natexlab{b}})}\BibitemShut {NoStop}%
%
%
\bibitem{PBF11}
V.-N. Phan, K.~W. Becker, and H. Fehske, EPL {\bf 95}, 17006 (2011).
\end{thebibliography}

%

\end{document}